\documentclass{iau} 
\usepackage{graphicx}

\newcommand{\chandra}{{\it Chandra~}}
\newcommand{\xmm}{{\it XMM-Newton~}}
\newcommand{\funit}{~erg cm$^{-2}$ s$^{-1}$}

\title[JD 11.~~Formation efficiency of HMXBs] 
{Different generations of HMXBs:\\clues about their formation efficiency\\from Magellanic Clouds studies}

\author[Vallia Antoniou et al. (on behalf of the SMC XVP collaboration)]   
{Vallia Antoniou$^{1,2}$, Andreas Zezas$^{3,4,1}$, Jeremy J. Drake$^1$, Carles Badenes$^5$, Frank Haberl$^6$, Jaesub Hong$^{1,7}$, Paul P. Plucinsky$^1$\\
 \and the SMC XVP Collaboration Team}

\affiliation{$^1$Harvard-Smithsonian Center for Astrophysics, Cambridge, MA, USA\\ email: {\tt vantoniou@cfa.harvard.edu} \\[\affilskip]
$^2$Texas Tech University, Department of Physics \& Astronomy, Lubbock, TX, USA\\[\affilskip]
$^3$University of Crete, Department of Physics, Heraklion, Greece\\[\affilskip]
$^4$IESL, Foundation for Research and Technology-Hellas, Heraklion, Greece\\[\affilskip]
$^5$University of Pittsburgh, Department of Physics \& Astronomy, Pittsburgh, PA, USA\\[\affilskip]
$^6$Max-Planck-Institut f{\"u}r extraterrestrische Physik, Garching, Germany\\[\affilskip]
$^7$Harvard University,Department of Astronomy, Cambridge, MA, USA 
}

\pubyear{2019}
\volume{346}  
\jname{High-Mass X-ray binaries: illuminating the passage from massive binaries to merging compact objects}
\editors{L.M. Oskinova, ed.}
\begin{document}

\maketitle

\begin{abstract}
Nearby star-forming galaxies offer a unique environment to study the populations of young ($<$100 Myr) accreting binaries. These systems are tracers of past populations of massive stars that heavily affect their immediate environment and parent galaxies. Using a Chandra X-ray Visionary program, we investigate the young neutron-star binary population in the low metallicity of the Small Magellanic Cloud (SMC) by reaching quiescent X-ray luminosity levels ($\sim$few times $10^{32}$ erg/s). We present the first measurement of the formation efficiency of high-mass X-ray binaries (HMXBs) as a function of the age of their parent stellar populations by using 3 indicators: the number ratio of HMXBs to OB stars, to the SFR, and to the stellar mass produced during the specific star-formation burst they are associated with. In all cases, we find that the HMXB formation efficiency increases as a function of time up to $\sim$40-60 Myr, and then gradually decreases.

\keywords{X-rays: binaries, (galaxies:) Magellanic Clouds, (stars:) pulsars: general, stars: neutron, stars: emission-line, Be, stars: early-type, stars: formation}
\end{abstract}

\firstsection 
\section{Introduction}

\begin{figure}[t]
\begin{center}
 \includegraphics[width=3.5in]{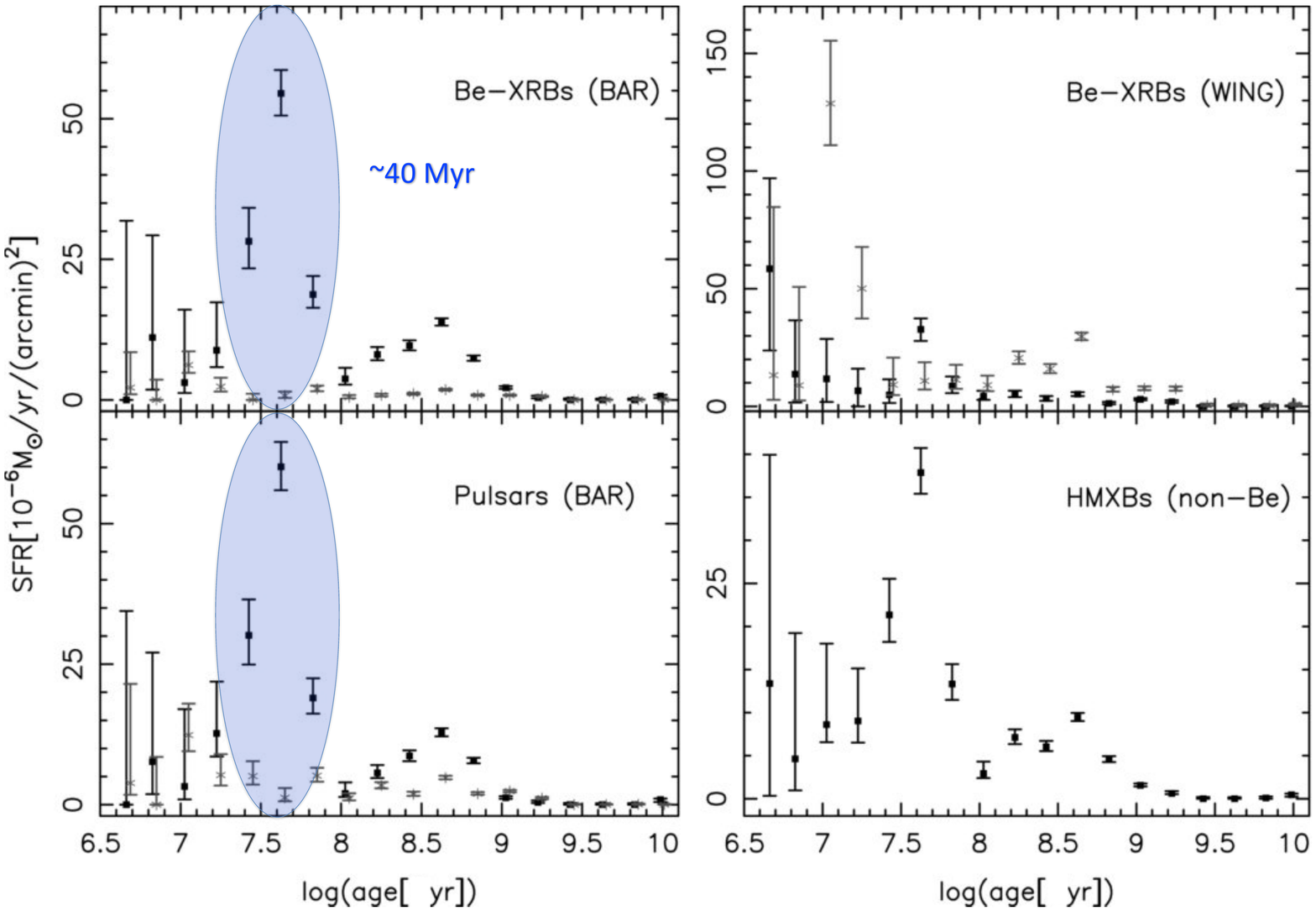} 
 \includegraphics[width=3.48in]{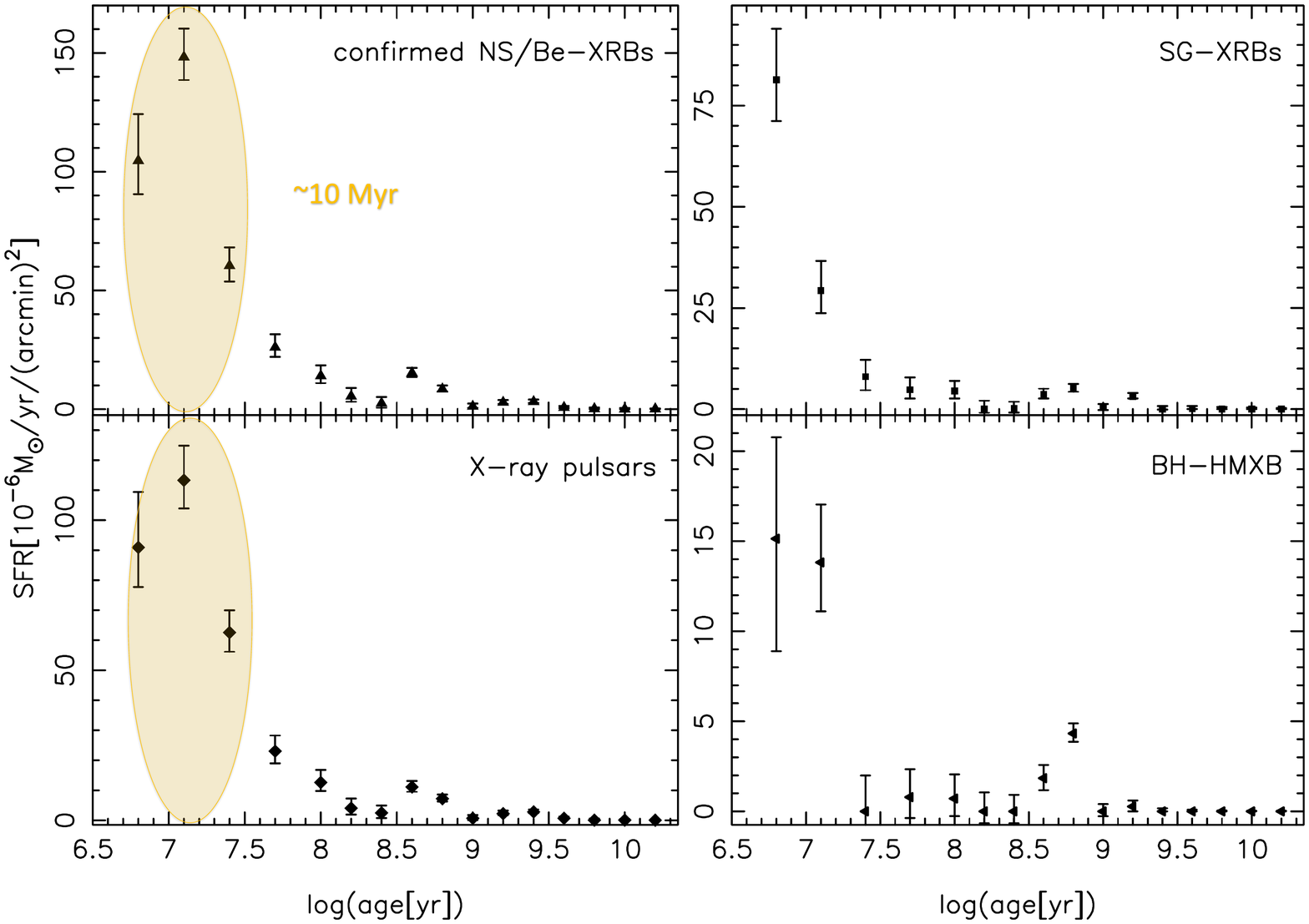} 
\caption{Average star-formation history of {\it (top)} SMC (\cite[Antoniou et al. 2010]{2010ApJ...716L.140A}) and {\it (bottom)} LMC (\cite[Antoniou \& Zezas 2016]{2016MNRAS.459..528A}) regions with (black) and without (grey) young X-ray binaries showing their association with stellar populations of different ages. A clear peak in the SF history of these regions appears at $\sim$40 Myr and $\sim$10 Myr, respectively.}\label{fig:SFH-MCs}
\end{center}
\end{figure}

Based on shallow \chandra and \xmm surveys of the Magellanic Clouds, we have found that the HMXBs, the Be/X-ray binaries (Be-XRBs; HMXBs with Oe- or Be-type companions), and the X-ray pulsars are observed in regions with star-formation rate (SFR) bursts $\sim$25--60 Myr (\cite[Antoniou et al. 2010]{2010ApJ...716L.140A}) and  $\sim$6--25 Myr (\cite[Antoniou \& Zezas 2016]{2016MNRAS.459..528A}) ago in the SMC and the LMC, respectively. In Fig.\,\ref{fig:SFH-MCs} we present the average star-formation history of regions in the Magellanic Clouds with and without young X-ray binaries showing their association with stellar populations of different ages.

\begin{figure}[t]
\begin{center}
 \includegraphics[width=2.625in]{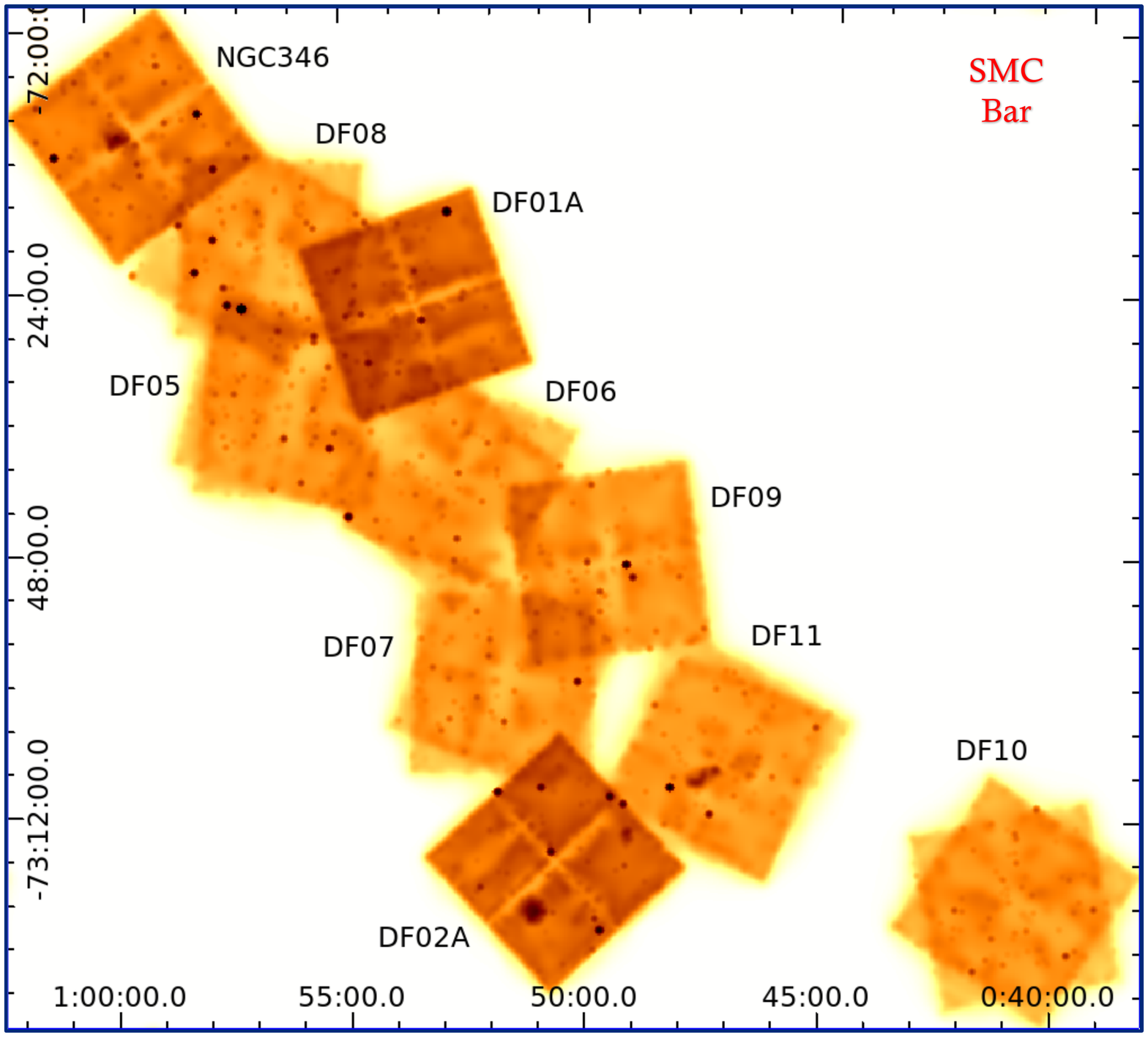} 
 \includegraphics[width=2.625in]{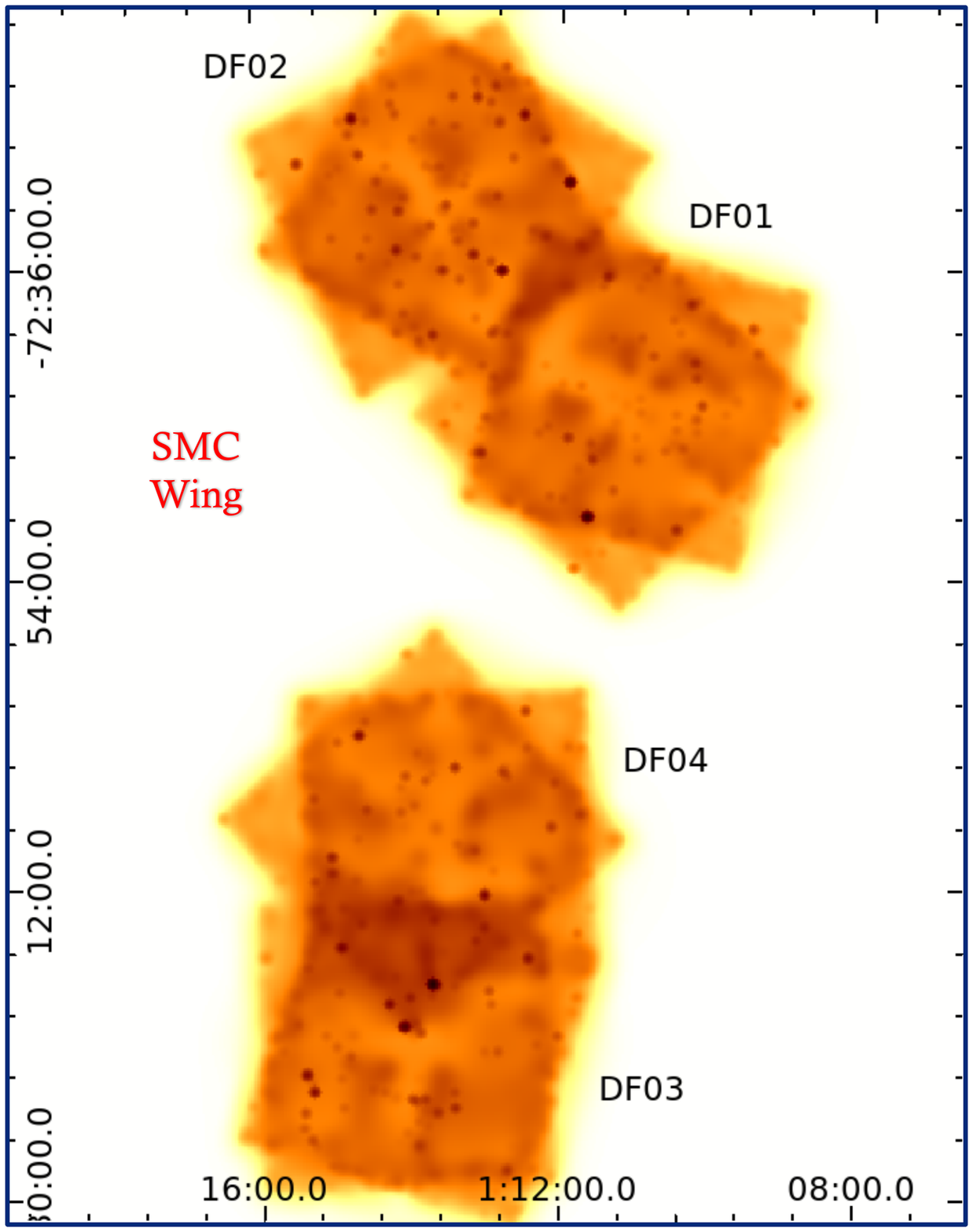} 
\caption{ACIS-I full band csmoothed exposure corrected images of the 11 Chandra X-ray Visionary fields analyzed in this work along with 3 archival exposures reaching the same depth.}\label{fig:ACIS-I}
\end{center}
\end{figure}

In order to study the young accreting binary population of the SMC in greater detail, we have been awarded a \chandra X-ray Visionary Program (XVP; PI A. Zezas) totaling 1.1 Ms of exposure time to observe  11 fields in this low-metallicity, star-forming galaxy. The fields were selected based on the age of their stellar populations, while on this dataset we have added 3 \chandra archival observations reaching the same 100 ks depth. The details of the analysis will be presented in Antoniou et al. (2019a, in prep.), while the timing analysis and its results have been already presented in  \cite[Hong et al. (2016)]{2016ApJ...826....4H} and \cite[Hong et al. (2017)]{2017ApJ...847...26H}. In total, we have detected 2,393 sources down to a limiting flux of $2.6 \times 10^{-16}$ \funit\, in the full (0.5 -- 8.0 keV) band (Fig.\,\ref{fig:ACIS-I}). 

\section{Identification of candidate HMXBs}
We then cross-correlated these X-ray sources with the OGLE-III optical photometric catalog of SMC stars (\cite[Udalski et al. 2008]{2008AcA....58..329U}), and placed their optical counterparts on a ($V, V-I$) color-magnitude diagram (CMD). Based on their positions with respect to the spectroscopically defined OB stars locus (Fig.\,\ref{fig:cparts}), we identified candidate HMXBs (following \cite[Antoniou et al. 2009a]{2009ApJ...697.1695A}, \cite[Antoniou et al. 2010]{2010ApJ...716L.140A}, and \cite[Antoniou \& Zezas 2016]{2016MNRAS.459..528A}). This approach has been particularly successful in identifying correct candidate members of the HMXB class of objects as proven by optical spectroscopic observations of candidate HMXBs (e.g. \cite[Antoniou et al. 2009b]{2009ApJ...707.1080A}), and H$\alpha$ narrow-band and $R$-band optical photometric surveys of the SMC (\cite[Maravelias et al. 2017]{2017IAUS..329..373M}; \cite[Maravelias et al. 2018]{2018arXiv181110933M}).

Next, we supplement the list of 127 candidate HMXBs identified in this work by 14 additional HMXBs identified by \cite[Haberl \& Sturm (2016)]{2016A&A...586A..81H} that fall within the XVP area but have not been detected in our survey or matched with any XVP source that has at least one OB counterpart in the OGLE-III catalog. The final list of HMXBs used in this work consists of 141 candidate and confirmed HMXBs.

\section{Age-dating of the HMXBs}
\begin{figure}[t]
\begin{center}
\includegraphics[width=2.85in,angle=270]{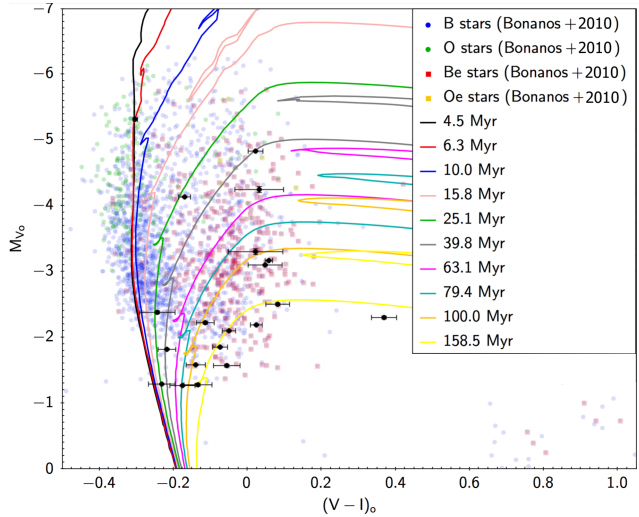} 
\caption{An example of the age determination of the HMXBs identified in \chandra field DF11 (black circles). $V$ and $V-I$ have been corrected for extinction as $M_{V_o} = (m-M)_{V} - A_{V} = V - 18.96 - 0.25$, and $(V-I)_{o} = (V-I) - E(V-I) = (V-I) - 0.13$.}\label{fig:cparts}
\end{center}
\end{figure}

In order to measure the HMXBs formation rate as a function of the age of their parent stellar populations, we first need to constrain the HMXBs ages and associate them with individual SF episodes responsible for the birth of their progenitors. Here, we estimate the age of the HMXBs from the optical counterpart positions on the ($V, V-I$) CMD with respect to the PARSEC isochrones (v1.2S; \cite[Bressan et al. 2012]{2012MNRAS.427..127B}) for ${\rm Z=0.004}$ (Fig.\,\ref{fig:cparts}).

Following that we associate each optical counterpart with a SF episode that overlaps with the age-range of the isochrones that are consistent with its location on the CMD. For example, out of the 17 HMXBs identified in \chandra field DF11 (Fig.\,\ref{fig:cparts}), only one has an optical counterpart photometrically consistent  with the SF peak at 7\,Myr, while the remaining 16 have ages consistent the second prominent SF peak at 42\,Myr.

\section{Formation efficiency of young accreting X-ray binaries}
\begin{figure}[t]
\begin{center}
 \includegraphics[width=6.5in]{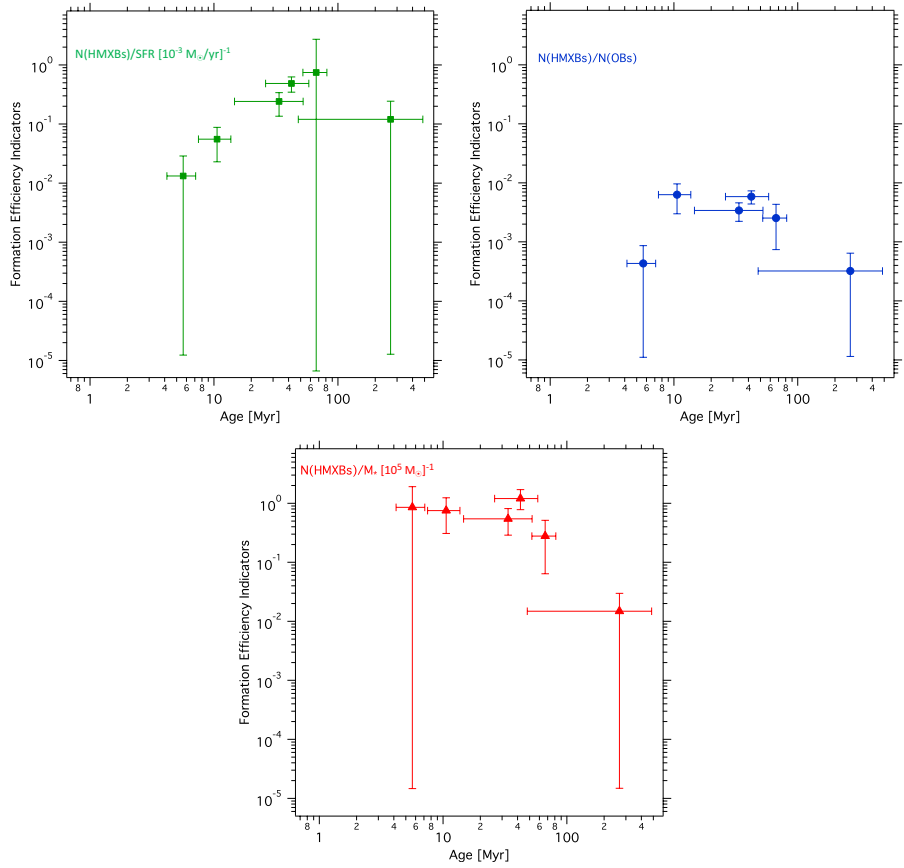} 
\caption{Different metrics of the HMXBs formation efficiency: number of HMXBs, N(HMXBs), with respect to the {\it (top left; green)} SFR of their parent stellar population; {\it (top right; blue)} number of OB stars, N(OBs); and {\it (bottom; red)} total stellar mass (M$\star$) formed during the SF episode they are associated with. All these indicators show an increase in the HMXBs formation rate for ages $\gtrsim$10--20\,Myr and up to 40--60\,Myr followed by a decline at older ages.}\label{fig:allFE} 
\end{center}
\end{figure}

We investigate three different indicators of the formation efficiency of HMXBs as a function of age: the number of HMXBs, N(HMXBs), in different ages with respect to: {\it (i)} the number of OB stars, N(OBs), in their respective \chandra field; {\it (ii)}
the SFR of their parent stellar population; and {\it (iii)} the total stellar mass (M$\star$) formed during the SF episode they are associated with.
The OB stars are from the OGLE-III catalog (\cite[Udalski et al. 2008]{2008AcA....58..329U}), while the SFRs as a function of age are from \cite[Harris \& Zaritsky (2004)]{2004AJ....127.1531H}. In Fig.\,\ref{fig:allFE}, we present {\it for the first time} these three indicators of the formation efficiency of HMXBs as a function of age. A more detailed analysis of the production rate of these systems is presented in Antoniou et al. (2019b, subm.).

The N(HMXBs)/N(OBs) ratio is an indicator that can be calculated directly for any nearby galaxy with resolved stellar populations, without the need to derive their SF history. Therefore, it serves as a useful proxy of the relative formation rate of HMXBs that can be applied to large samples of galaxies. It also takes into account the present-day numbers of OB stars. On the other hand, N(HMXBs)/SFR takes into account the SF event that created the binaries we observe today, but not the duration of the SF burst. Finally, the ratio N(HMXBs)/M$\star$ takes into account the SF burst duration (M$\star$ is the integral of the SFR as a function of time), and is the fundamental relation that we were aiming to derive from this \chandra XVP program. We also note that the delay function of the HMXBs (Antoniou et al. 2019b, subm.) resembles best the ratio N(HMXBs)/M$\star$.

In all three cases, we find an increase in the formation rate of the HMXBs for ages $\gtrsim$10--20\,Myr and up to 40--60\,Myr followed by a decline at older ages. 
This result is in agreement with studies of the formation efficiency of massive Oe/Be stars in the Magellanic Clouds (e.g. \cite[Bonanos et al. 2009]{2009AJ....138.1003B}, \cite[Bonanos et al. 2010]{2010AJ....140..416B}), and the Milky Way (\cite[McSwain \& Gies 2005]{2005ApJS..161..118M}) that show a peak at ages of $\sim$20--50\,Myr, matching the age of maximum production of HMXBs in the SMC.

\vspace{0.5cm}
VA acknowledges financial support from NASA/Chandra grants GO3-14051X, AR4-15003X, NNX15AR30G, NASA/ADAP grant NNX10AH47G and the Texas Tech President's Office. AZ acknowledges financial support from NASA/ADAP grant NNX12AN05G and funding from the European Research Council under the European Union's Seventh Framework Programme (FP/2007-2013)/ERC Grant Agreement n.~617001. This project has also received funding from the European Union's Horizon 2020 research and innovation programme  under the Marie Sklodowska-Curie RISE action,  grant agreement No 691164 (ASTROSTAT). JJD and PPP were funded by NASA contract NAS8-03060 to the {\it Chandra X-ray Center}.

\end{document}